\documentclass[apjl]{emulateapj}
\usepackage{epsfig}
\usepackage{dcolumn}
\usepackage{bm}
\usepackage{natbib}
\usepackage{times}
\usepackage{ulem}
\def\kmm#1  {{\bf [KMM:~ #1]~}}
\def\new#1 {{\bf #1 }}
\def\cut#1 {\sout{#1} }

\newcommand{\dal}{\ensuremath{\lsb \Delta \alpha/ \alpha \rsb}}

\newcommand{\dmu}{\ensuremath{\lsb \Delta \mu/\mu \rsb}}
\newcommand{\beq}{\begin{equation}}
\newcommand{\eeq}{\end{equation}}

\newcommand{\lb}{\left(}
\newcommand{\rb}{\right)}
\newcommand{\lsb}{\left[}
\newcommand{\rsb}{\right]}

\shorttitle{Constraining changes in the proton-electron mass ratio}
\shortauthors{Kanekar}
\begin{document}
\title{Constraining changes in the proton-electron mass ratio with inversion and rotational lines}

\author{Nissim Kanekar\altaffilmark{1}}
\altaffiltext{1}{Ramanujan Fellow, National Centre for Radio Astrophysics, 
TIFR, Ganeshkhind, Pune - 411007, India}

\begin{abstract}
We report deep Green Bank Telescope (GBT) spectroscopy in the redshifted NH$_3$~(1,1),
CS~1-0 and H$_2$CO~0$_{\rm 00}$-1$_{\rm 01}$ lines from the $z \sim 0.685$ absorber towards
B0218+357. The inversion (NH$_3$) and rotational (CS, H$_2$CO) line frequencies have 
different dependences on the proton-electron mass ratio $\mu$, implying that a comparison 
between the line redshifts is sensitive to changes in $\mu$. A joint 3-component fit 
to the NH$_3$, CS, and H$_2$CO lines yields $\dmu = (-3.5 \pm 1.2) \times 10^{-7}$,
from $z \sim 0.685$ to today, where the error includes systematic effects from 
comparing lines from different species and possible frequency-dependent source 
morphology. Two additional sources of systematic error remain, due to time variability in 
the source morphology and velocity offsets between nitrogen-bearing and carbon-bearing 
species. We find no statistically-significant ($\ge 3 \sigma$) evidence for changes in 
$\mu$, and obtain the stringent $3\sigma$ constraint, $\dmu < 3.6 \times 10^{-7}$, over 
6.2~Gyrs; this is the best present limit on temporal changes in $\mu$ from any technique, 
and for any lookback time, by a factor $\gtrsim 5$.

\end{abstract}


\section{Introduction} 
\label{sec:intro}

A generic prediction of theories that attempt to unify the standard model of particle 
physics with general relativity is that particle masses and low-energy coupling constants 
vary with space and time. A detection of such changes would imply new physics beyond 
the standard model. Tests of changes in fundamental ``constants'' such as the fine 
structure constant $\alpha$, the proton-electron mass ratio $\mu$ or the proton 
g-factor $g_p$ are hence of much interest (e.g. \citealp{uzan03}).

Redshifted atomic and molecular spectral lines provide an interesting avenue to 
probe the possibility of fundamental constant evolution over cosmological timescales 
\citep{savedoff56}. Comparisons between the redshifts of lines whose rest frequencies 
have different dependences on a constant like $\alpha$ are sensitive to changes 
in the constant [see \citet{kanekar08b} for a recent review]. There are a number 
of such techniques, using comparisons between different transitions (e.g. 
\citealp{thompson75,wolfe76,dzuba99,darling03,chengalur03,kanekar04a,flambaum07b}). 
Indeed, one of these techniques, the many-multiplet method, has yielded 
evidence for changes in $\alpha$: \citet{murphy04} obtained $\dal = (-5.4 \pm 1.2) 
\times 10^{-6}$ from 143~absorbers with $\langle z \rangle = 1.75$ (see also 
\citealp{murphy03}). This result has as yet been neither confirmed nor ruled out 
(e.g. \citealp{srianand07b,molaro08,murphy08b,kanekar10}), but evidence has now 
been found for additional systematic effects in the data, due to errors in the 
wavelength calibration \citep{griest10}. Recently, an independent technique, 
using radio ``conjugate'' satellite OH lines \citep{kanekar04b,kanekar05alt}, found 
weak evidence (at $\sim 99.1$\% confidence level) for changes in a combination of $\alpha$, 
$\mu$ and $g_p$: \citet{kanekar10b} obtained $\left[\Delta G/G \right] = (-1.18 \pm 0.46) 
\times 10^{-5}$ between $z \sim 0.247$ and the present epoch, where 
$G \equiv g_p [ \mu \alpha^2]^{1.85}$. If changes in $\alpha$ are assumed to dominate 
over those in $\mu$ and $g_p$, one obtains $\dal = (-3.1 \pm 1.2) \times 10^{-6}$, 
consistent with the result of \citet{murphy04}, albeit at a lower redshift.

For three decades, ro-vibrational molecular hydrogen (H$_2$) lines provided the sole method
to directly probe changes in the proton-electron mass ratio \citep{thompson75}. Unfortunately,
few redshifted H$_2$ absorbers have been detected so far (e.g. \citealt{ledoux03}), 
and, of these, only four have been found suitable to study changes in $\mu$ 
(e.g. \citealt{ivanchik05,reinhold06,king08,thompson09,malec10,wendt11}). These results 
have also been controversial: \citet{reinhold06} obtained $\dmu = (+2.0 \pm 0.6) \times 10^{-5}$
from two absorbers at $z \sim 2.6 - 3.0$, i.e. evidence for a larger value of 
$\mu$ at high redshift (see also \citealt{ivanchik05}). However, 
independent re-analyses of these data by \citet{king08} and \citet{thompson09}, 
using improved wavelength calibration techniques, found no evidence for changes in 
$\mu$. \citet{king08} obtain $\dmu = (+2.6 \pm 3.0) \times 10^{-6}$ from three absorbers 
at $z \sim 2.6 - 3.0$ (including the two systems of \citealt{reinhold06}); note, however,
that \citet{wendt11} argue that systematic errors have been under-estimated in
the latter analysis.

A new laboratory technique to probe changes in $\mu$ was proposed by \citet{vanveldhoven04},
using inversion transitions of deuterated ammonia (ND$_3$). \citet{flambaum07b} 
adapted this technique to astrophysical circumstances, using ammonia (NH$_3$) inversion 
lines and rotational lines. Rotational and inversion line frequencies have different 
dependences on $\mu$, and a comparison between the redshifts of NH$_3$ inversion lines 
and rotational (e.g. CO, HCO$^+$, CS, etc) lines from a single cosmologically-distant 
absorber is thus sensitive to changes in $\mu$. Only two redshifted NH$_3$ absorbers 
are currently known, at $z \sim 0.685$ towards B0218+357 \citep{henkel05} and $z \sim 0.886$ 
towards B1830$-$210 \citep{henkel08}. In both cases, the NH$_3$ detection spectra 
have been used to obtain initial constraints on changes in $\mu$ \citep{murphy08,henkel09}. 
Finally, the ammonia technique has also been used to search for spatial changes in $\mu$: 
\citet{levshakov09} obtained the (conservative) constraint $\dmu < 3 \times 10^{-8}$ 
from studies of multiple molecular clouds in the Galaxy.

The $z \sim 0.685$ absorber towards B0218+357 has a far smaller velocity spread 
than the $z \sim 0.886$ system towards B1830$-$210 \citep{wiklind95,wiklind96b}, 
making it a better candidate for accurate redshift measurements. The carbon monosulfide 
(CS)~1-0 line [rest frequency: 48.990957(2)~GHz; \citealp{kim03}] is a good rotational 
transition for the comparison with the NH$_3$ inversion lines because it is 
known to be unsaturated in the $z \sim 0.685$ absorber \citep{combes97b} and it 
lies at a frequency relatively close to that of the NH$_3$~(1,1) lines. 
The formaldehyde (H$_2$CO)~0$_{\rm 00}$-1$_{\rm 01}$ line [rest frequency: $72.837948 (10)$~GHz; 
\citealp{cornet80}] was chosen as a second rotational transition, to obtain an estimate 
of systematic effects in the above comparison; this too is known to be unsaturated in 
the $z \sim 0.685$ absorber \citep{jethava07}. 
This {\it Letter} reports Green Bank Telescope (GBT) spectroscopy in 
the redshifted NH$_3$~(1,1), CS~1-0 and H$_2$CO~0$_{\rm 00}$-1$_{\rm 01}$ lines, that
yield the best present constraints on changes in the proton-electron mass ratio.

\section{Observations, data analysis and results}
\label{sec:spectra}

\begin{figure*}[t!]
\includegraphics[scale=0.4]{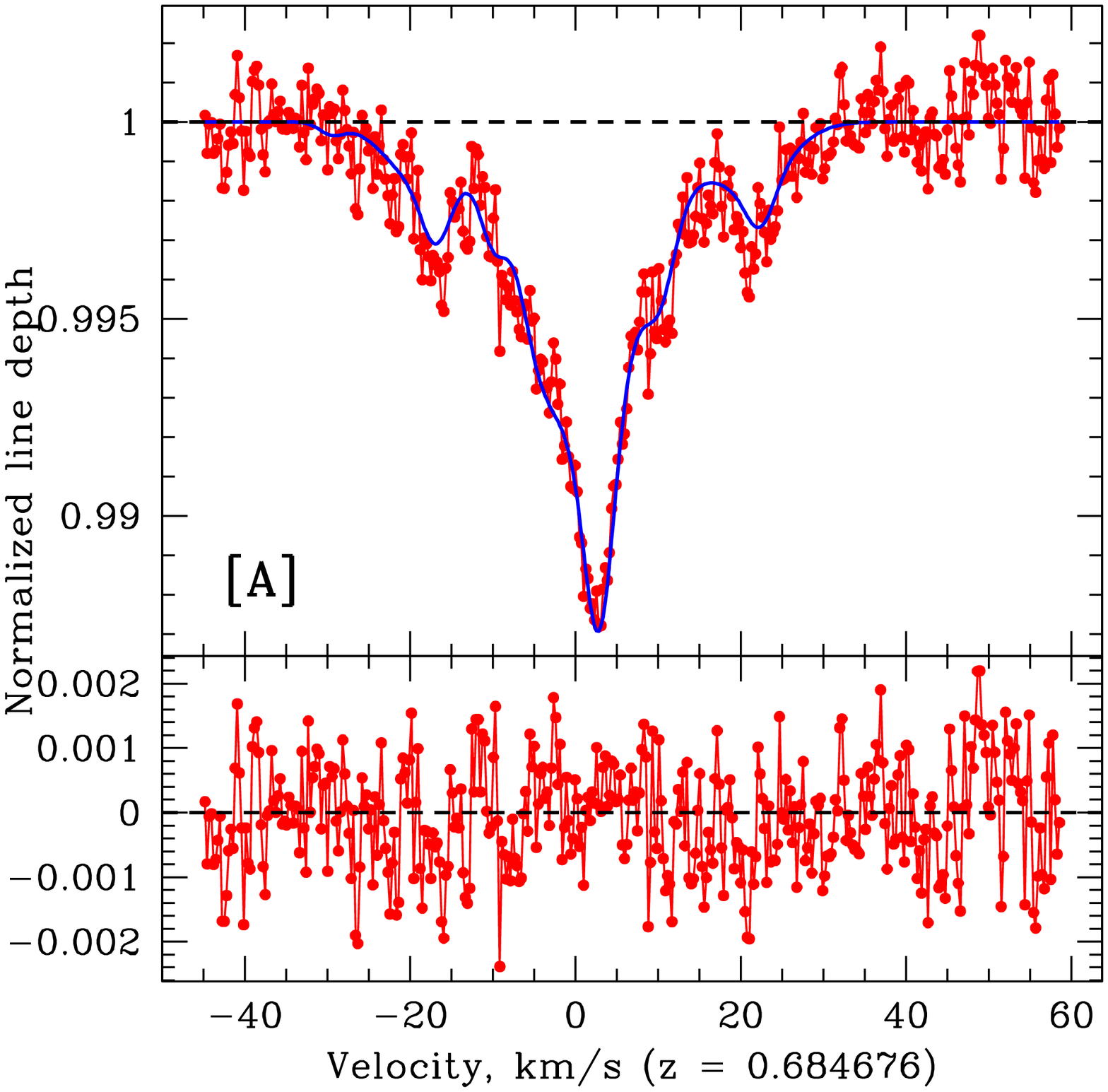}
\includegraphics[scale=0.4]{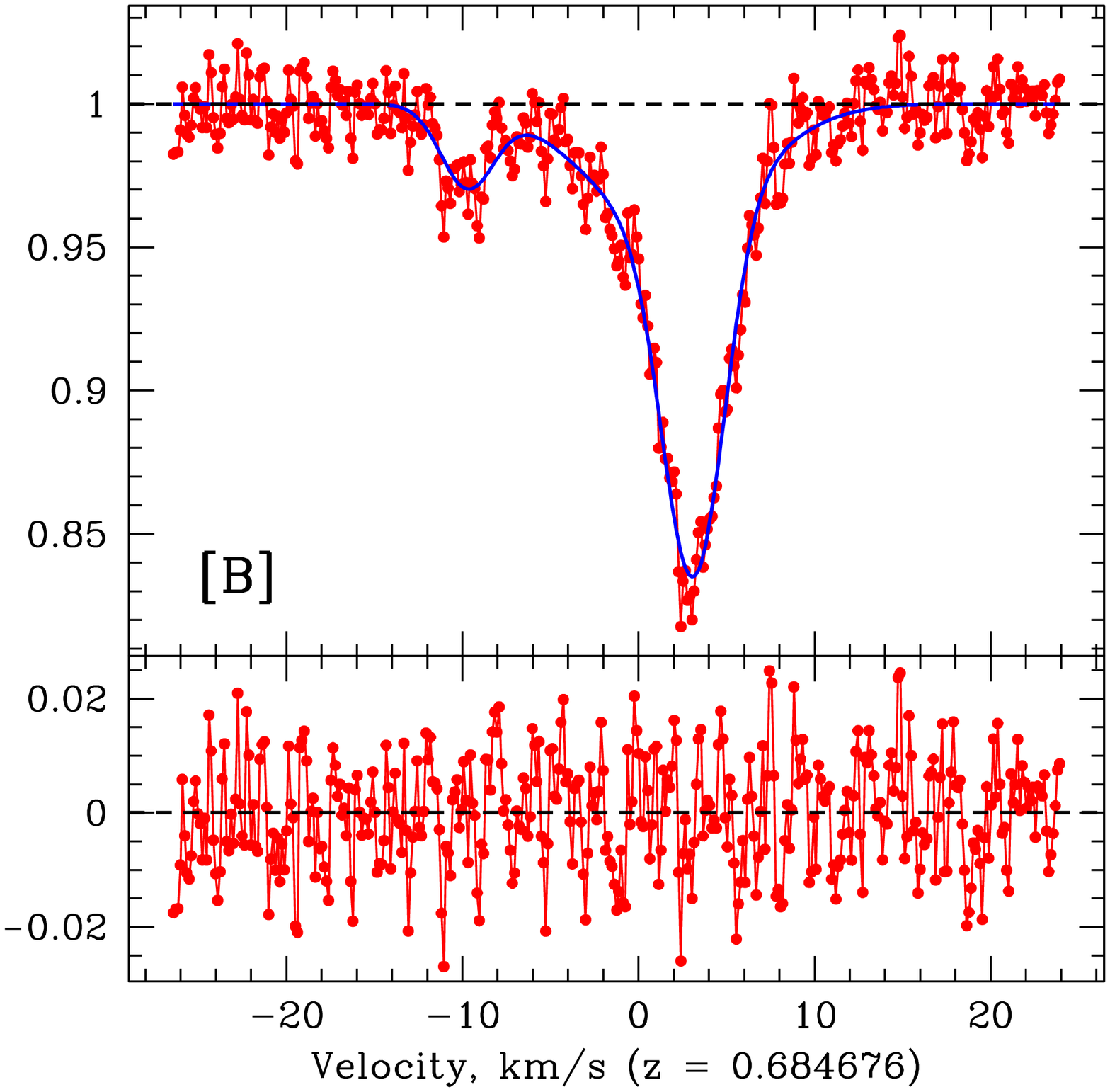}
\includegraphics[scale=0.4]{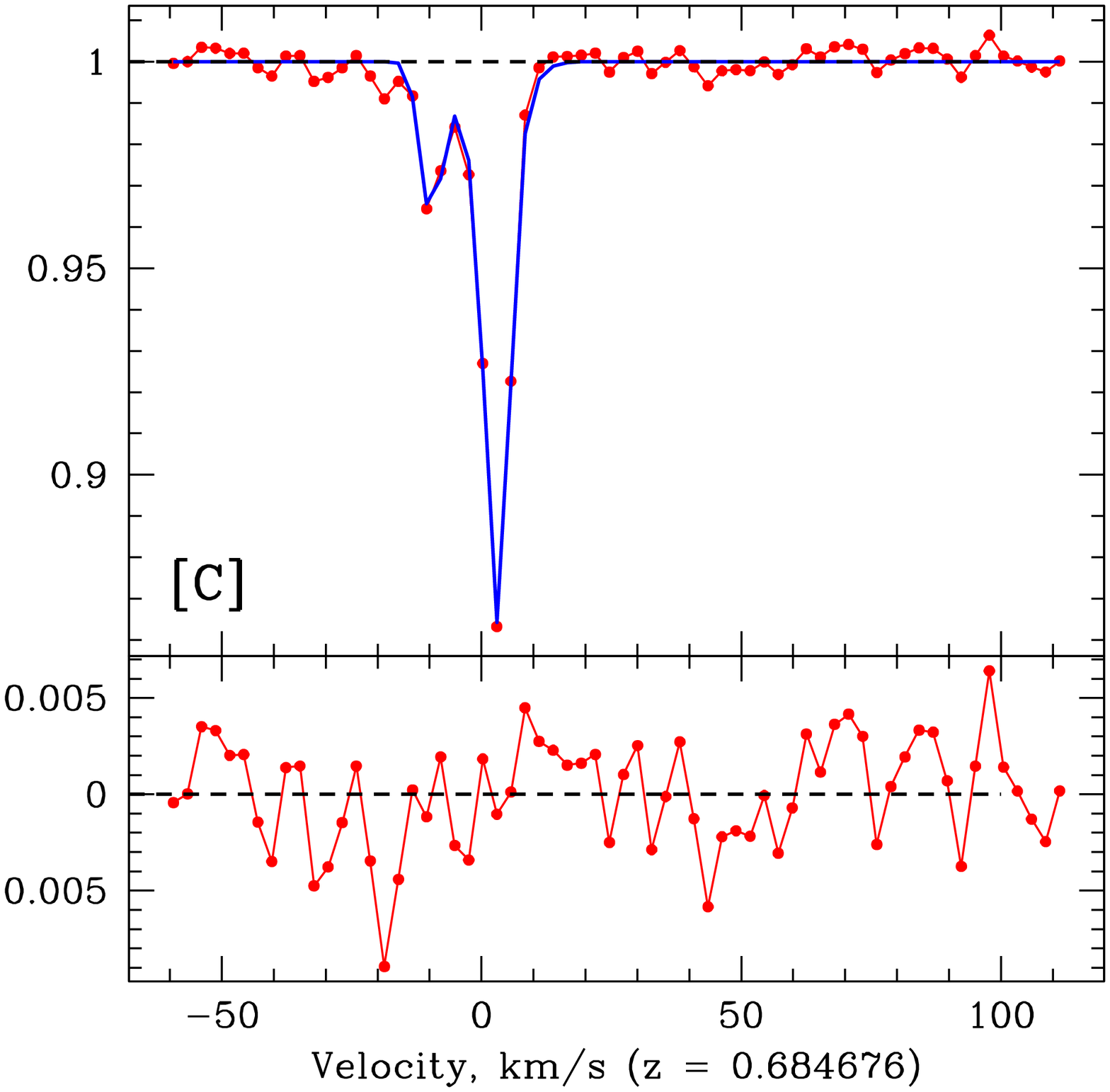}
\caption{GBT spectra in the redshifted [A]~NH$_3$~(1,1), [B]~CS~1-0 and 
[C]~H$_2$CO~0$_{\rm 00}$-1$_{\rm 01}$ transitions from the $z \sim 0.685$ 
absorber towards B0218+357. The upper panels show normalized line depth in 
each transition plotted against velocity, in km/s, relative to a heliocentric 
redshift of $z = 0.684676$; the 3-component fit is overlaid on each spectrum. The lower 
panels show the residuals from the spectra after subtracting out the joint 3-component fit; 
these are consistent with noise.  \label{fig:fig1}}
\end{figure*}

The GBT observations of the NH$_3$~(1,1) and CS~1-0 lines at $z \sim 0.685$ towards 
B0218+357 were carried out in January and August 2008 (proposals AGBT07C-016 and 
AGBT07C-054), using the Ku- and Ka-band receivers, respectively. A spectrum in 
the redshifted H$_2$CO~0$_{\rm 00}$-1$_{\rm 01}$ line was obtained in July~2008 
with the GBT Q-band receiver,
as part of an absorption survey of the $z \sim 0.685$ galaxy (proposal AGBT07C-054). 
Bandpass calibration was carried out via beam-switching for the NH$_3$ observations, 
and with sub-reflector nodding for the higher-frequency CS and H$_2$CO observations. 
Two polarizations were used for the observations of the NH$_3$ (Ku-band) and H$_2$CO 
(Q-band) lines, and a single Ka-band polarization for the CS line. The total 
observing times were 6~hours (NH$_3$), 1.5~hours (CS) and 3~hours (H$_2$CO), 
with the NH$_3$ observations split into two observing sessions, in January and August 2008. 
Standard procedures in the 
packages {\sc DISH} (for the NH$_3$ lines) and {\sc GBTIDL} (for the CS and H$_2$CO lines) 
were used to analyse the data. The final NH$_3$, CS, and H$_2$CO spectra are shown 
in Figs.~\ref{fig:fig1}[A], [B] and [C]; these have velocity resolutions of $\sim 0.26$, 
$0.12$ and $2.8$~km/s, respectively (after Hanning-smoothing and re-sampling), and 
root-mean-square noise values of $\sim 0.00082$, $0.0097$ and $0.0026$, in 
optical depth units, per independent velocity channel.

A GBT Ku-band search was also carried out for the CCS~1-0 transition at $z \sim 0.685$;
no absorption was detected at the redshifted CCS line frequency, with a $3\sigma$ optical 
depth limit of $\sim 0.0016$ per 4.2~km/s channel. The K- and Ku-band receivers were 
further used to search for redshifted NH$_3$~(1,1) absorption from two other 
redshifted molecular absorbers, at $z \sim 0.247$ towards B1413+135 \citep{wiklind97} 
and $z \sim 0.674$ towards B1504+377 \citep{wiklind96}, with no detected 
absorption in either case. The $3\sigma$ optical depth limits were 
$0.00068$~per 4~km/s channel (B1413+135) and $0.00061$ per 4~km/s channel
(B1504+377).

The test for changes in the proton-electron mass ratio $\mu$ was carried out 
through a simultaneous fit to the NH$_3$~(1,1), CS~1-0 and 
H$_2$CO~0$_{\rm 00}$-1$_{\rm 01}$ spectra, using the package {\sc VPFIT}\footnote{http://www.ast.cam.ac.uk/\~rfc/vpfit.html}.
The velocity structure in all optically-thin lines was assumed to be the same, 
in both the number of absorbing ``clouds'' and their redshifts. Turbulence was 
assumed to be the dominant contributor to the line widths, which were hence tied 
together in the fit \citep{murphy08}. The NH$_3$ hyperfine structure was included 
by assuming local thermodynamic equilibrium (LTE), with the hyperfine ratios taken 
from Table~S1 of \citep{murphy08}. Finally, the fit also included a single velocity 
offset between the spectral components in the inversion and rotational transitions, 
to account for a possible change in the proton-electron mass ratio. 

\setcounter{table}{0}
\begin{table*}
\begin{centering}
\begin{tabular}{|c|c|c|c|c|c|c|}
\hline
&&&&\multicolumn{3}{c|}{} \\ 
Component & Velocity offset &  Redshift & FWHM   & \multicolumn{3}{c|}{Line depth $\times 100$} \\
 	  &    km/s         &           &  km/s  &  NH$_3$ & CS & H$_2$CO \\
\hline
&&&&&& \\ 
1 & & $0.6846935 (2)$ & $3.99 \pm 0.13$ & $0.496 \pm 0.060$ & $13.83 \pm 0.78$ & $15.61 \pm 0.84$ \\ 
2 & $(-0.36 \pm 0.10$) & $0.6846858 (10)$ & $10.60 \pm 0.45$ & $0.734 \pm 0.026$ & $4.39  \pm 0.33$ & $3.00 \pm 0.39$ \\ 
3 & & $0.6846214 (7)$ & $3.53 \pm 0.33$ & $0.128 \pm 0.032$ & $2.84  \pm 0.28$ & $5.17 \pm 0.37$ \\ 
&&&&&& \\ 
\hline
\end{tabular}
\caption{Parameters of the best 3-component Voigt profile fit to the NH$_3$~(1,1), 
CS~1-0 and H$_2$CO~0$_{\rm 00}$-1$_{\rm 01}$ spectra. The second column contains the 
best-fit velocity offset between the rotational and inversion lines; note that a 
single velocity offset was assumed for all spectral components. The redshifts of 
column~(3) are in the heliocentric frame, while the line depths ($\times 100$), listed 
in the last three columns, have been normalized by the source continuum at each 
observing frequency. \label{table:fit} }
\end{centering}
\end{table*}

A simultaneous multi-component Voigt profile fit was carried out to the three spectra 
of Fig.~\ref{fig:fig1} with the above assumptions, aiming to minimize 
$\chi^2_\nu$ by varying the fit parameters. A 3-component fit was found to yield 
$\chi^2_\nu = 1.05$ and no evidence for structure in the residuals after subtracting out 
the fits to each spectrum (see the lower panels of Fig.~\ref{fig:fig1}). Specifically, 
a Kolmogorov-Smirnov rank-1 test found the fit residuals for all lines to be 
consistent with a normal distribution within $\sim 1 \sigma$ confidence. One- and 
two-component fits were found to yield both a larger $\chi^2_\nu$ and clear 
non-Gaussian structure in the residual spectra. While increasing the number of 
spectral components to four did yield a marginally-lower $\chi^2_\nu$ ($\sim 1.02$), 
the additional spectral component was only weakly detected in the H$_2$CO spectrum 
(at $\lesssim 3\sigma$ significance) and was not visible in the NH$_3$ or CS transitions. 
A 3-component model thus appears to provide a good fit to the data; however,
for completeness, results from the 4-component fit will also be mentioned below.

For the coupled 3-component model, the number of fit parameters is 16, three redshifts,
three line widths, nine optical depths in the NH$_3$, CS, and H$_2$CO lines, and
the velocity offset between the inversion and rotational transitions. The parameters 
of the best 3-component fit are listed in Table~1; the best-fit velocity offset 
is $\Delta V = (-0.36 \pm 0.10)$~km/s, with the NH$_3$ lines blueshifted relative 
to the rotational lines. A similar result is obtained from the best-fit 
4-component model, $\Delta V = (-0.39 \pm 0.11)$~km/s. Similar results are 
also obtained from carrying out 3-component fits to the NH$_3$ and CS lines 
(i.e. without including the H$_2$CO line), and to the NH$_3$ and H$_2$CO lines
(i.e. without the CS line).

To obtain an estimate of systematic effects, a similar 3-component fit was 
also carried out to the CS and H$_2$CO rotational lines alone (i.e. {\it without} 
including the NH$_3$ inversion lines). This yielded a velocity offset of 
$\delta V = (0.029 \pm 0.068)$~km/s, between the CS and H$_2$CO lines. The $1\sigma$
error obtained here, when comparing two rotational lines, will be used as an
estimate of the systematic error due to local velocity offsets between different 
species in the absorber.

The fractional change in the proton-electron mass ratio $\mu$ is related to the 
measured velocity offset between inversion and rotational lines $\Delta V$ by
the expression \citep{flambaum07b} $\dmu = 
0.289 \lsb {\lb z_{inv} - z_{rot} \rb}/{\lb 1+\bar{z}\rb}\rsb 
\approx 0.289 \left[{\Delta V}/{c}\right]$, 
where $z_{inv}$ and $z_{rot}$ are, respectively, the inversion and rotational 
redshifts, and $\bar{z}$ is their average. Our final velocity offset is 
$\Delta V = [-0.36 \pm 0.10 (stat.) \pm 0.068 (syst.)]$~km/s; this yields 
$\dmu = [-3.47 \pm 0.96 (stat.) \pm 0.66 (syst.)] \times 10^{-7}$. Adding 
the statistical and systematic errors in quadrature, we obtain $\dmu = 
(-3.5 \pm 1.2) \times 10^{-7}$, between $z \sim 0.685$ and the present epoch.

\section{Discussion}
\label{sec:discussion}

An excellent summary of the systematic effects inherent in the comparison between 
rotational and inversion lines in the $z \sim 0.685$ absorber is given by 
\citet{murphy08}. These include the following: (1)~the assumption 
that different transitions have the same number of spectral components, and at the 
same redshifts, (2)~the background source morphology is frequency-dependent, implying 
that the NH$_3$ and rotational transitions might arise along slightly different 
sightlines, (3)~the background source flux density (and hence morphology) varies 
with time, so observations at different epochs might probe different sightlines, 
(4)~the assumption of LTE for the NH$_3$ hyperfine structure, (5)~saturation effects 
if highly-saturated lines [e.g. the HCO$^+$ and HCN lines used by \citet{murphy08}] 
are used, and (6)~local velocity offsets between the species giving rise to the 
transitions. As noted by \citet{murphy08}, all these effects could yield significant 
contributions to their systematic errors. For example, the inversion and rotational 
transitions used by \citet{murphy08} were at very different frequencies, $\sim 14$~GHz 
and $\sim 106$~GHz; changes in the background source morphology could thus result in 
different sightlines in the different transitions. The NH$_3$ 
and HCO$^+$/HCN observations were separated by a few years, implying that temporal
variability in the source morphology could also be an issue, especially given 
that saturated transitions were used in the analysis. Finally, the NH$_3$ spectra 
were of too low signal-to-noise ratio (S/N) to detect the NH$_3$ hyperfine structure 
and to test for non-LTE effects.

Most of the above issues have been directly addressed in the present work. 
All three transitions used in the analysis are unsaturated and have been 
observed at high S/N, and with the same telescope. The two outlying NH$_3$ 
hyperfine components (at $\pm 19.5$~km/s relative to the strongest component) 
in the NH$_3$~(1,1) line have also been clearly detected (see Fig.~\ref{fig:fig1}[A]). 
As noted by \citet{murphy08}, non-LTE effects resulting in hyperfine ``anomalies'' 
should cause the satellite hyperfine components (with $\Delta F_1 \ne 0$, where $F_1$ 
is the quadrupole quantum number) to have different optical depths. No evidence for 
such non-LTE conditions is apparent in the $\pm 19.5$~km/s satellite hyperfine 
components in Fig.~\ref{fig:fig1}[A]; the optical depths are found to agree 
within the noise.

We note, finally, that the systematic error of $7.6 \times 10^{-7}$ in the result 
of \citet{murphy08} contains two contributions: (1)~$7 \times 10^{-7}$ because 
spectral components detected in the HCO$^+$ and HCN spectra may not be 
detected in the NH$_3$ spectra, due either to the large frequency difference between 
the inversion and rotation lines or to the fact that the HCO$^+$ and HCN lines are 
optically thick, and (2)~$3 \times 10^{-7}$ due to the possibility of non-LTE 
effects in the NH$_3$ spectra. Our choice of rotational lines and the higher
sensitivity of our NH$_3$ spectra implies that neither of these are significant 
sources of systematic error in the present analysis.

\citet{liszt06} find that NH$_3$ column densities correlate best with CS and 
H$_2$CO column densities in Galactic diffuse clouds; CS and H$_2$CO are thus likely 
to be the best rotational transitions for the inversion/rotation comparison, as the 
correlation in column densities suggests that the three species are likely to 
arise in the same part of a gas cloud. Galactic absorption has also been
detected in all three species towards two quasars, B0355+508 and B0415+379 
\citep{liszt95,lucas98,liszt06}; both sightlines show two components in all species, 
with line velocities agreeing within 0.2~km/s [see Table~1 of \citealt{liszt95}, 
Table~6 of \citealt{lucas98} and Table~A1 of \citealt{liszt06}]; local velocity
offsets between the species thus appear to be small. Further, the rotational 
and inversion line frequencies are much closer here than in the comparison of 
\citet{murphy08}, with the redshifted NH$_3$, CS, and H$_2$CO lines at $\sim 14$~GHz, 
$\sim 29$~GHz and $\sim 43$~GHz, respectively. Changes in source morphology with frequency 
are thus a less important issue here than in the analysis of \citet{murphy08}. A direct 
test comes from the comparison between the two rotational lines, whose frequencies differ 
by a factor of $\sim 1.5$, comparable to the ratio of the CS and NH$_3$ frequencies ($\sim 2$). 
The CS-H$_2$CO comparison also provides an estimate of the systematic error due to 
local velocity offsets between different species in the clouds. This test found the 
CS and  H$_2$CO line redshifts to agree within the noise; the $1\sigma$ error in this
comparison has been used to quantify the systematic error due to both local velocity
offsets and differing background source morphology at the different line frequencies.
Finally, similar velocity offsets (consistent within the errors) were obtained in 
the independent comparisons between the NH$_3$ and CS lines, and the NH$_3$ and 
H$_2$CO lines.

A source of systematic effects that could not be directly addressed here is 
time variability in the background source morphology, which might yield different 
sightlines through the absorbing clouds at different epochs [\citealp{murphy08}; 
see \citet{muller08} for the $z \sim 0.886$ lens towards B1830$-$210]. Unfortunately, 
the redshifted NH$_3$, CS and H$_2$CO transitions towards B0218+357 require different 
GBT receivers and cannot be observed simultaneously.  The present NH$_3$, CS and 
H$_2$CO observations were carried out between January and August~2008; the possibility 
that the weak offset between the inversion and rotational lines arises due to 
small changes in the sightline thus cannot be ruled out. Further, the agreement between 
CS and H$_2$CO velocities does not rule out the possibility that nitrogen-bearing 
species like NH$_3$ arise at different velocities than carbon-bearing species like 
CS and H$_2$CO. This can only be tested by using rotational transitions of other 
nitrogen-bearing species (e.g. HC$_3$N, CH$_3$CN, etc).

While there are two possible sources of systematic error that we are as yet unable 
to quantify, the present data show no statistically-significant ($\ge 3\sigma$) 
evidence for changes in the proton-electron mass ratio. Our $3\sigma$ upper limit on 
changes in $\mu$ is $\dmu < 3.6 \times 10^{-7}$, between $z \sim 0.685$ and $z = 0$ 
(a lookback time of 6.2~Gyrs). For comparison, \citet{murphy08} obtained 
$\dmu < 1.8 \times 10^{-6}$ ($2\sigma$) in the $z \sim 0.685$ 
absorber towards B0218+357, while \citet{henkel09} obtained $\dmu < 1.4 \times 10^{-6}$ 
($3\sigma$) in the $z \sim 0.886$ absorber towards B1830$-$210, with both results based 
on inversion/rotation comparisons. Note, however, that \citet{henkel09} used 
single-Gaussian fits for the NH$_3$ and HC$_3$N lines towards B1830$-$210, although 
it is clear from the profiles of other unsaturated lines (e.g. CS) that 
at least three absorbing ``clouds'' are present along the sightline.  Further, the 
GBT spectra in the strongest [(1,1), (2,2) and (3,3)] NH$_3$ lines towards B1830$-$210 
were severely affected by radio frequency interference (see Fig.~1 of \citealt{henkel08}).
The effect of these issues on the results of \citet{henkel09} is unclear, but the 
error budget is likely to increase. At higher redshifts, the best published constraint 
on changes in $\mu$ is that of \citet{king08}: $\dmu < 6.0 \times 10^{-6}$, using H$_2$ 
lines (but see \citealt{wendt11}). The present result, $\dmu < 3.6 \times 10^{-7}$ ($3\sigma$) 
at $z \sim 0.685$, is thus the most sensitive constraint on temporal changes in 
$\mu$ at any redshift, by a factor $\gtrsim 5$. 

To compare this result with those from laboratory studies, it is necessary to assume a model 
for the variation of $\mu$ with time. For linear variation, the present result yields 
$\dot{\mu}/\mu < 5.6 \times 10^{-17}$ per year, more than an order of magnitude 
better than the best (model-dependent) laboratory constraint on changes in $\mu$ 
\citep{rosenband08alt}.

In conclusion, we have used the GBT to obtain high S/N spectra in the 
NH$_3$~(1,1), CS~1-0 and H$_2$CO~0$_{\rm 00}$-1$_{\rm 01}$ transitions from the $z \sim 0.685$
absorber towards B0218+357. A comparison between the redshifts of the inversion and 
rotational lines yields a velocity offset of $\Delta V = [-0.36 \pm 0.10 (stat.) 
\pm 0.068 (syst.)]$~km/s, with the NH$_3$ lines blueshifted relative to the rotational 
ones. Two sources of systematic error remain to be quantified, arising from 
(1)~velocity offsets between nitrogen-bearing and carbon-bearing species, and 
(2)~time variability in the background source morphology, which might yield different 
sightlines at different epochs. We find no statistically-significant ($\ge 3\sigma$) 
evidence for changes in $\mu$, with the $3\sigma$ constraint $\dmu < 3.6 \times 
10^{-7}$ over a lookback time of 6.2~Gyrs. This is the strongest present constraint 
on temporal changes in the proton-electron mass ratio.

\acknowledgments
I thank Bob Carswell and Michael Murphy for much help with using {\sc VPFIT} for 
the Voigt-profile fitting,  Carl Bignell and Bob Garwood for help in the scheduling 
of the GBT observations and the GBT data analysis, and Dave Meier for permission to use 
the H$_2$CO line detected in our survey of B0218+357 for this project. I also
acknowledge support from the Department of Science and Technology, India, via a 
Ramanujan Fellowship. The National Radio Astronomy Observatory is operated by 
Associated Universities, Inc, under cooperative agreement with the NSF. 




\begin{thebibliography}{}

\bibitem[{Chengalur \& Kanekar(2003)}]{chengalur03}
Chengalur, J.~N. \& Kanekar, N. 2003, Phys.~Rev.~Lett., 91, 241302

\bibitem[{{Combes} {et~al.}(1997){Combes}, {Wiklind}, \& {Nakai}}]{combes97b}
{Combes}, F., {Wiklind}, T., \& {Nakai}, N. 1997, A\&A, 327, L17

\bibitem[{{Cornet} \& {Winnewisser}(1980)}]{cornet80}
{Cornet}, R.~A. \& {Winnewisser}, G. 1980, J.Mol.Spec., 80, 438

\bibitem[{Darling(2003)}]{darling03}
Darling, J. 2003, Phys.~Rev.~Lett., 91, 011301

\bibitem[{Dzuba {et~al.}(1999)Dzuba, Flambaum, \& Webb}]{dzuba99}
Dzuba, V.~A., Flambaum, V.~V., \& Webb, J.~K. 1999, Phys.~Rev.~Lett., 82, 888

\bibitem[{{Flambaum} \& {Kozlov}(2007)}]{flambaum07b}
{Flambaum}, V.~V. \& {Kozlov}, M.~G. 2007, Phys.~Rev.~Lett., 98, 240801

\bibitem[{{Griest} {et~al.}(2010){Griest}, {Whitmore}, {Wolfe}, {Prochaska},
  {Howk}, \& {Marcy}}]{griest10}
{Griest}, K., {Whitmore}, J.~B., {Wolfe}, A.~M., {Prochaska}, J.~X., {Howk},
  J.~C., \& {Marcy}, G.~W. 2010, ApJ, 708, 158

\bibitem[{{Henkel} {et~al.}(2008){Henkel}, {Braatz}, {Menten}, \&
  {Ott}}]{henkel08}
{Henkel}, C., {Braatz}, J.~A., {Menten}, K.~M., \& {Ott}, J. 2008, A\&A, 485,
  451

\bibitem[{{Henkel} {et~al.}(2005){Henkel}, {Jethava}, {Kraus}, {Menten},
  {Carilli}, {Grasshoff}, {Lubowich}, \& {Reid}}]{henkel05}
{Henkel}, C., {Jethava}, N., {Kraus}, A., {Menten}, K.~M., {Carilli}, C.~L.,
  {Grasshoff}, M., {Lubowich}, D., \& {Reid}, M.~J. 2005, A\&A, 440, 893

\bibitem[{{Henkel} {et~al.}(2009){Henkel}, {Menten}, {Murphy}, {Jethava},
  {Flambaum}, {Braatz}, {Muller}, {Ott}, \& {Mao}}]{henkel09}
{Henkel}, C., {Menten}, K.~M., {Murphy}, M.~T., {Jethava}, N., {Flambaum},
  V.~V., {Braatz}, J.~A., {Muller}, S., {Ott}, J., \& {Mao}, R.~Q. 2009, A\&A,
  500, 725

\bibitem[{{Ivanchik} {et~al.}(2005){Ivanchik}, {Petitjean}, {Varshalovich},
  {Aracil}, {Srianand}, {Chand}, {Ledoux}, \& {Boiss{\'e}}}]{ivanchik05}
{Ivanchik}, A., {Petitjean}, P., {Varshalovich}, D., {Aracil}, B., {Srianand},
  R., {Chand}, H., {Ledoux}, C., \& {Boiss{\'e}}, P. 2005, A\&A, 440, 45

\bibitem[{{Jethava} {et~al.}(2007){Jethava}, {Henkel}, {Menten}, {Carilli}, \&
  {Reid}}]{jethava07}
{Jethava}, N., {Henkel}, C., {Menten}, K.~M., {Carilli}, C.~L., \& {Reid},
  M.~J. 2007, A\&A, 472, 435

\bibitem[{{Kanekar}(2008)}]{kanekar08b}
{Kanekar}, N. 2008, Mod. Phys. Lett. A, 23, 2711

\bibitem[{{Kanekar} \& {Chengalur}(2004)}]{kanekar04a}
{Kanekar}, N. \& {Chengalur}, J.~N. 2004, MNRAS, 350, L17

\bibitem[{{Kanekar} {et~al.}(2004){Kanekar}, {Chengalur}, \&
  {Ghosh}}]{kanekar04b}
{Kanekar}, N., {Chengalur}, J.~N., \& {Ghosh}, T. 2004, Phys. Rev. Lett., 93, 051302

\bibitem[{{Kanekar} {et~al.}(2010{\natexlab{a}}){Kanekar}, {Chengalur}, \&
  {Ghosh}}]{kanekar10b}
---. 2010{\natexlab{a}}, ApJ, 716, L23

\bibitem[{{Kanekar et al.}(2005)}]{kanekar05alt}
{Kanekar}, N. {et al.} 2005, Phys.~Rev.~Lett., 95, 261301

\bibitem[{{Kanekar} {et~al.}(2010{\natexlab{b}}){Kanekar}, {Prochaska},
  {Ellison}, \& {Chengalur}}]{kanekar10}
{Kanekar}, N., {Prochaska}, J.~X., {Ellison}, S.~L., \& {Chengalur}, J.~N.
  2010{\natexlab{b}}, ApJ, 712, L148

\bibitem[{{Kim} \& {Yamamoto}(2003)}]{kim03}
{Kim}, E. \& {Yamamoto}, S. 2003, J.Mol.Spec., 219, 296

\bibitem[{{King} {et~al.}(2008){King}, {Webb}, {Murphy}, \&
  {Carswell}}]{king08}
{King}, J.~A., {Webb}, J.~K., {Murphy}, M.~T., \& {Carswell}, R.~F. 2008, Phys.
  Rev. Lett., 101, 251304

\bibitem[{{Ledoux} {et~al.}(2003){Ledoux}, {Petitjean}, \&
  {Srianand}}]{ledoux03}
{Ledoux}, C., {Petitjean}, P., \& {Srianand}, R. 2003, MNRAS, 346, 209

\bibitem[{{Levshakov} {et~al.}(2010){Levshakov}, {Molaro}, {Lapinov},
  {Reimers}, {Henkel}, \& {Sakai}}]{levshakov09}
{Levshakov}, S.~A., {Molaro}, P., {Lapinov}, A.~V., {Reimers}, D., {Henkel},
  C., \& {Sakai}, T. 2010, A\&A, 512, 44

\bibitem[{Liszt \& Lucas(1995)}]{liszt95}
Liszt, H. \& Lucas, R. 1995, A\&A, 299, 847

\bibitem[{{Liszt} {et~al.}(2006){Liszt}, {Lucas}, \& {Pety}}]{liszt06}
{Liszt}, H., {Lucas}, R., \& {Pety}, J. 2006, A\&A, 448, 253

\bibitem[{{Lucas} \& {Liszt}(1998)}]{lucas98}
{Lucas}, R. \& {Liszt}, H. 1998, A\&A, 337, 246

\bibitem[{{Malec} {et~al.}(2010){Malec}, {Buning}, {Murphy}, {Milutinovic},
  {Ellison}, {Prochaska}, {Kaper}, {Tumlinson}, {Carswell}, \&
  {Ubachs}}]{malec10}
{Malec}, A.~L., {Buning}, R., {Murphy}, M.~T., {Milutinovic}, N., {Ellison},
  S.~L., {Prochaska}, J.~X., {Kaper}, L., {Tumlinson}, J., {Carswell}, R.~F.,
  \& {Ubachs}, W. 2010, MNRAS, 403, 1541

\bibitem[{{Molaro} {et~al.}(2008){Molaro}, {Reimers}, {Agafonova}, \&
  {Levshakov}}]{molaro08}
{Molaro}, P., {Reimers}, D., {Agafonova}, I.~I., \& {Levshakov}, S.~A. 2008,
  European Physical Journal Special Topics, 163, 173

\bibitem[{{Muller} \& {Gu{\'e}lin}(2008)}]{muller08}
{Muller}, S. \& {Gu{\'e}lin}, M. 2008, A\&A, 491, 739

\bibitem[{{Murphy} {et~al.}(2008{\natexlab{a}}){Murphy}, {Flambaum}, {Muller},
  \& {Henkel}}]{murphy08}
{Murphy}, M.~T., {Flambaum}, V.~V., {Muller}, S., \& {Henkel}, C.
  2008{\natexlab{a}}, Science, 320, 1611

\bibitem[{{Murphy} {et~al.}(2004){Murphy}, {Flambaum}, {Webb}, {Dzuba},
  {Prochaska}, \& {Wolfe}}]{murphy04}
{Murphy}, M.~T., {Flambaum}, V.~V., {Webb}, J.~K., {Dzuba}, V.~V., {Prochaska},
  J.~X., \& {Wolfe}, A.~M. 2004, in Lecture Notes in Physics, Vol. 648,
  Astrophysics, Clocks and Fundamental Constants, ed. S.~G. {Karshenboim} \&
  E.~{Peik} (Berlin: Springer-Verlag), 131

\bibitem[{Murphy {et~al.}(2003)Murphy, Webb, \& Flambaum}]{murphy03}
Murphy, M.~T., Webb, J.~K., \& Flambaum, V.~V. 2003, MNRAS, 345, 609

\bibitem[{{Murphy} {et~al.}(2008{\natexlab{b}}){Murphy}, {Webb}, \&
  {Flambaum}}]{murphy08b}
{Murphy}, M.~T., {Webb}, J.~K., \& {Flambaum}, V.~V. 2008{\natexlab{b}}, MNRAS,
  384, 1053

\bibitem[{{Reinhold} {et~al.}(2006){Reinhold}, {Buning}, {Hollenstein},
  {Ivanchik}, {Petitjean}, \& {Ubachs}}]{reinhold06}
{Reinhold}, E., {Buning}, R., {Hollenstein}, U., {Ivanchik}, A., {Petitjean},
  P., \& {Ubachs}, W. 2006, Phys.~Rev.~Lett., 96, 151101

\bibitem[{{Rosenband et al.}(2008)}]{rosenband08alt}
{Rosenband}, T. {et al.} 2008, Science, 319, 1808

\bibitem[{{Savedoff}(1956)}]{savedoff56}
{Savedoff}, M.~P. 1956, Nature, 178, 688

\bibitem[{{Srianand} {et~al.}(2007){Srianand}, {Chand}, {Petitjean}, \&
  {Aracil}}]{srianand07b}
{Srianand}, R., {Chand}, H., {Petitjean}, P., \& {Aracil}, B. 2007, Phys. Rev. Lett., 
 99, 239002

\bibitem[{{Thompson}(1975)}]{thompson75}
{Thompson}, R.~I. 1975, ApL, 16, 3

\bibitem[{{Thompson} {et~al.}(2009){Thompson}, {Bechtold}, {Black},
  {Eisenstein}, {Fan}, {Kennicutt}, {Martins}, {Prochaska}, \&
  {Shirley}}]{thompson09}
{Thompson}, R.~I., {Bechtold}, J., {Black}, J.~H., {Eisenstein}, D., {Fan}, X.,
  {Kennicutt}, R.~C., {Martins}, C., {Prochaska}, J.~X., \& {Shirley}, Y.~L.
  2009, ApJ, 703, 1648

\bibitem[{Uzan(2003)}]{uzan03}
Uzan, J.-P. 2003, Rev. Mod. Phys, 75, 403

\bibitem[{{van Veldhoven} {et~al.}(2004){van Veldhoven}, {K{\"u}pper},
  {Bethlem}, {Sartakov}, {van Roij}, \& {Meijer}}]{vanveldhoven04}
{van Veldhoven}, J., {K{\"u}pper}, J., {Bethlem}, H.~L., {Sartakov}, B., {van
  Roij}, A.~J.~A., \& {Meijer}, G. 2004, Eur. Phys. Jour. D, 31, 337

\bibitem[{{Wendt} \& {Molaro}(2011)}]{wendt11}
{Wendt}, M. \& {Molaro}, P. 2010, A\&A, 526, 96

\bibitem[{{Wiklind} \& {Combes}(1995)}]{wiklind95}
{Wiklind}, T. \& {Combes}, F. 1995, A\&A, 299, 382

\bibitem[{{Wiklind} \& {Combes}(1996{\natexlab{a}})}]{wiklind96}
---. 1996{\natexlab{a}}, A\&A, 315, 86

\bibitem[{{Wiklind} \& {Combes}(1996{\natexlab{b}})}]{wiklind96b}
---. 1996{\natexlab{b}}, \nat, 379, 139

\bibitem[{{Wiklind} \& {Combes}(1997)}]{wiklind97}
---. 1997, A\&A, 328, 48

\bibitem[{{Wolfe} {et~al.}(1976){Wolfe}, {Broderick}, {Condon}, \&
  {Johnston}}]{wolfe76}
{Wolfe}, A.~M., {Broderick}, J.~J., {Condon}, J.~J., \& {Johnston}, K.~J. 1976,
  ApJ, 208, L47

\end{thebibliography}
\end{document}